\documentstyle[prb,aps,eqsecnum,epsf,preprint]{revtex}

\begin{document}
\draft
\title{Dimensional crossover and metal-insulator transition in 
quasi-two-dimensional disordered conductors  }
\author{ N. Dupuis\thanks{On leave from Laboratoire de Physique des Solides,
Universit\' e Paris-Sud,
91405 Orsay, France }}
\address{Department of Physics, University of Maryland, College Park, MD
20742-4111} 
\maketitle
\begin{abstract}
We study the metal-insulator transition (MIT) in weakly coupled disordered
planes on the basis of a Non-Linear Sigma Model (NL$\sigma $M). Using two
different methods, a renormalization group (RG) approach and an auxiliary
field method, we calculate
the crossover length between a 2D regime at small length scales and a 3D
regime at larger length scales. The 3D regime is described by an anisotropic
3D NL$\sigma $M with renormalized coupling constants. We obtain the critical
value of the single particle interplane hopping which separates the metallic
and insulating phases. We also show that a strong parallel magnetic field
favors the localized phase and derive the phase diagram.  
\end{abstract}

\section{Introduction}

A quasi-two-dimensional (quasi-2D) weakly disordered conductor (i.e. weakly
coupled planes system) exhibits a MIT for a critical value of the 
single particle interplane hopping $t_\perp $. The existence of
this quantum phase transition results from the localization of the
electronic states in a 2D system by arbitrary weak disorder, while a 3D
system remains metallic below a critical value of the disorder. 

Although the existence of this MIT is well established, its properties are
far from being understood. In particular, the critical value $t_\perp
^{(c)}$ of the interplane coupling remains controversial. The
self-consistent diagrammatic theory of the Anderson localization
\cite{Vollhardt92}  
predicts an exponentially small critical coupling \cite{Prigodin84} 
$t_\perp ^{(c)}\sim \tau ^{-1}e^{-\alpha k_Fl}$ ($\alpha \sim 1$) in
agreement with simple scaling arguments \cite{ND92} and other analytical
arguments.\cite{Li89} (Here $\tau $ is the elastic scattering time, $k_F$
the 2D Fermi wave vector and $l$ the intraplane mean free path. 
$k_Fl\gg 1$ for weak
disorder.) A recent numerical analysis predicts a completely
different result, $t_\perp ^{(c)}\sim 1/\sqrt{\tau }$, which is supported by
analytical arguments. \cite{Zambetaki96} It has also been claimed, on the
basis of diagrammatic perturbative calculations, that the MIT depends on
the propagating direction, in contradiction with the scaling theory of
localization. \cite{Abrikosov94} (Note that for a system of weakly coupled
chains, it is well established, both from numerical and analytical
approaches, that $t_\perp ^{(c)}\sim 1/\tau $. \cite{Dorokhov83})

The aim of this paper is to reconsider the MIT in quasi-2D conductors on the
basis of a NL$\sigma $M.\cite{Belitz94} In the next
section, we use a renormalization group (RG) approach to show how the system
crosses over from a 2D behavior at small length scales to a 3D behavior at
larger length scales. The crossover length $L_x$ is explicitly
calculated. We show that the 3D regime is described by an effective
NL$\sigma $M with renormalized coupling constants. 
We obtain the critical value of
the interplane coupling and the anisotropy of the correlation (localization)
lengths in the metallic (insulating) phase. Next, we show that a parallel
magnetic field tends to decouple the planes and thus can induce a MIT. The
results of the RG approach are recovered in section \ref{sec:AF} by means of
an auxiliary field method. We emphasize that the latter is a general
approach to study phase transitions in weakly coupled systems.

\section{Renormalization Group approach }

We consider spinless electrons propagating in a quasi-2D
system with the dispersion law 
\begin{equation}
\epsilon_{\bf k}={\bf k}_\parallel ^2/2m -2t_\perp \cos (k_\perp d) \,,
\end{equation}
where ${\bf k}_\parallel $ and $k_\perp $ are the longitudinal
(i.e. parallel to the planes) and transverse components of ${\bf k}$,
respectively. 
$m$ is the effective mass in the planes, $d$ the interplane spacing and
$t_\perp $ the transfer integral in the transverse ($z$) direction.
The effect of disorder is taken into account by adding a random potential
with zero mean and gaussian probability distribution:
\begin{equation}
\langle V_l({\bf r}) V_{l'}({\bf r}') \rangle =(2\pi N_2(0)\tau )^{-1}
\delta _{l,l'}\delta ({\bf r}-{\bf r}') \,,
\end{equation}
where $N_2(0)=m/2\pi $ is the 2D density of states at the Fermi level and $\tau
$ the elastic scattering time. ${\bf r}$ is the coordinate in the plane and the
integers $l,l'$ label the different planes. We note $k_F$ the 2D Fermi wave
vector and $v_F=k_F/m$ the 2D Fermi velocity.  

\subsection{A simple argument}

We first determine the critical value $t_\perp ^{(c)}$ by means of a
simple argument whose validity will be confirmed in the next sections.   
Consider
an electron in a given plane. If the coupling $t_\perp $ is sufficiently
weak, the electron will first diffuse in the plane and then hop to the
neighboring plane after a time $\tau _x$. The corresponding diffusion length
$L_x$ is determined by the equations
\begin{eqnarray}
L_x^2&=&D(L_x) \tau _x \,, \nonumber \\
d^2&=&D_\perp \tau _x=2t_\perp ^2d^2\tau \tau _x \,.
\label{harg}
\end{eqnarray}
The length dependence of the coefficient $D(L)$ results from quantum
corrections to the semiclassical 2D diffusion coefficient $D=v_F^2\tau /2$. 
We have assumed
that the transverse diffusion is correctly described by the semiclassical
diffusion coefficient $D_\perp $. As shown in the next sections, this is
a consequence of the vanishing scaling dimension of the field in the
NL$\sigma $M approach. Eqs.\ (\ref{harg}) give $\tau _x\sim 1/t_\perp ^2\tau
$ and $L_x^2\sim D(L_x)/t_\perp ^2\tau $. The critical value $t_\perp ^{(c)}$
is obtained from the condition $L_x\sim \xi _{2D}$ or $g(L_x)\sim
1$, where $\xi _{2D}$ is the 2D localization length and $g(L)$ the 2D
(dimensionless) conductance. These two conditions are equivalent since
$g(L)=N_2(0)D(L)\sim 
1$ for $L\sim \xi _{2D}$ (see Eq.\ (\ref{g2D}) below). We thus obtain  
\begin{equation}
t_\perp ^{(c)} \sim \frac{1}{\sqrt{\xi _{2D}^2N_2(0)\tau }} 
\sim \frac{l}{\xi _{2D}\sqrt{k_Fl}}\frac{1}{\tau }  \,,
\label{tcrit}
\end{equation}
where $l=v_F\tau $ is the elastic mean free path. In the weak disorder limit
($k_Fl\gg 1$),
$\xi _{2D}\sim le^{\alpha k_Fl}$ ($\alpha \sim 1$) so that
$t_\perp ^{(c)}$ is exponentially small with respect to $1/\tau $. Apart
from the factor $1/\sqrt{k_Fl}$, Eq.\ (\ref{tcrit}) agrees with the result
of the self consistent diagrammatic theory of Anderson
localization\cite{Prigodin84} and with estimates based on the weak
localization correction to the Drude-Boltzmann conductivity. \cite{WL}

\subsection{NL$\sigma $M for weakly coupled planes}

The procedure to derive the NL$\sigma $M describing electrons in a random
potential is well established \cite{Belitz94} and we only quote the final
result in the quasi-2D case (see however section \ref{subsec:MF}). Averaging
over disorder by introducing $N$ replica of the system and using an
imaginary time formalism, the effective action of the model is $S_\Lambda
=S_{2D}+S_\perp $ with    
\begin{eqnarray}
S_{2D} \lbrack Q \rbrack &=& {\pi \over 8}N_2(0)D \sum _l \int d^2r {\rm
Tr} \lbrack \bbox{ \nabla }_\parallel 
Q_l({\bf r})\rbrack ^2 - {\pi \over
2}N_2(0) \sum _l \int d^2r 
{\rm Tr}\lbrack \Omega Q_l({\bf r})\rbrack \,,  \nonumber \\     
S_\perp \lbrack Q \rbrack &=& - {\pi \over 4}N_2(0)t_\perp ^2\tau \sum
_{\langle l,l' \rangle } \int d^2r 
{\rm Tr} \lbrack Q_l({\bf r}) Q_{l'}({\bf r})\rbrack  \,,
\label{action1}
\end{eqnarray}
where $Q_l({\bf r})$ is a matrix field with elements 
$_{ij}Q_{lnm}^{\alpha \beta }({\bf r})$. $\alpha ,\beta =1...N$ are
replica indices, $n,m$ refer to fermionic Matsubara frequencies $\omega
_n,\omega _m$, and $i,j=1,2$ describe the particle-hole ($i=j$) and
particle-particle ($i\neq j$) channels.
$\Omega $ is the matrix $_{ij}\Omega _{lnm}^{\alpha \beta }=
\delta _{i,j}\delta _{n,m}\delta _{\alpha ,\beta }\omega _n$. 
In (\ref{action1}), Tr denotes the trace over all discrete
indices. The field $Q$ satisfies the constraints $Q_l^2({\bf r})=\underline
1$ (with $\underline 1$ the unit matrix), ${\rm
Tr}\, Q_l({\bf r})=0$ and $Q^+=C^TQ^TC=Q$ where $C$ is the charge conjugaison
operator. \cite{Belitz94} $\Lambda \sim 1/l$ is an ultraviolet cut-off for
the fluctuations of the field $Q$ in the planes. The Josephson-like
interplane coupling $S_\perp $ is obtained retaining the
lowest order contribution in $t_\perp $ ($\langle l,l' \rangle $ denotes
nearest neighbors).     

If $t_\perp ^2\tau \gg D\Lambda ^2\sim 1/\tau $, the fluctuations in the
transverse direction are weak. In this case, ${\rm Tr}\lbrack
Q_lQ_{l'}\rbrack $ can be approximated by $-(d^2/2){\rm Tr}\lbrack \nabla
_zQ\rbrack ^2$  and we obtain a NL$\sigma
$M with anisotropic diffusion coefficients $D$ and $D_\perp $. For $t_\perp
\tau \gg 1$, an electron in a given plane hops to the neighboring plane
before being scattered. There is therefore no 2D diffusive motion in that
limit so that the quasi-2D aspect does not play an essential role. The
anisotropy can be simply eliminated by an appropriate  length rescaling in the
longitudinal and transverse directions. \cite{Wolfle84,possible} 
We consider in the following only the limit $t_\perp \tau \ll 1$ where we
expect a 2D/3D dimensional crossover at a characteristic length $L_x$.

\subsection{RG approach}

We analyze the action (\ref{action1}) within a RG approach following Ref.\
\onlinecite{Affleck}. The condition $t_\perp \tau \ll 1$ ensures that the
transverse coupling is weak ($D\Lambda ^2\gg t_\perp ^2\tau $). The initial
stage of the renormalization will therefore be essentially 2D, apart from
small corrections due to the interplane coupling. If we neglect the latter, 
we then obtain the renormalized action
\begin{eqnarray}
S_{\Lambda '} \lbrack Q \rbrack &=& {\pi \over 8}N_2(0)D(\Lambda ') 
\sum _l \int d^2r {\rm Tr} \lbrack \bbox{ 
\nabla }_\parallel  Q_l({\bf r})\rbrack ^2 - {\pi \over 2}N_2(0) \sum _l
\int d^2r 
{\rm Tr}\lbrack \Omega Q_l({\bf r})\rbrack  \nonumber \\     
&& - {\pi \over 4}N_2(0)t_\perp ^2\tau \sum _{\langle l,l' \rangle } \int d^2r
{\rm Tr} \lbrack Q_l({\bf r}) Q_{l'}({\bf r})\rbrack \,,
\label{action2}
\end{eqnarray}
where $\Lambda '<\Lambda $ is the reduced cut-off after renormalization and
$D(\Lambda ')$ the renormalized value of the 2D longitudinal diffusion 
coefficient.  
Since the scaling dimension of the field $Q$ vanishes (i.e. there is no
rescaling of the field), \cite{Belitz94} there is no renormalization of the
interplane coupling. \cite{Affleck} The 2D/3D dimensional crossover occurs
when the transverse and longitudinal couplings become of the same order: 
\begin{equation}
{1 \over 2}D_x\Lambda _x^2 \sim t_\perp ^2\tau \,, 
\label{critXover}
\end{equation}
where $D_x=D(\Lambda _x)$ is the longitudinal diffusion coefficient at the
crossover. In the 3D regime ($\Lambda '\leq \Lambda _x$), it is appropriate to
take the continuum limit in the transverse direction.\cite{Affleck}  Using 
\begin{equation}
{\rm Tr}\lbrack Q_lQ_{l'}\rbrack  = -{1 \over 2} {\rm Tr}\lbrack Q_l-Q_{l'}
\rbrack ^2 +{\rm const}
\to -\frac{d^2}{2} {\rm Tr}\lbrack \nabla _zQ_l\rbrack ^2 
\end{equation}
for $l$ and $l'$ nearest neighbors, and $d\sum _l\to \int dz $, we obtain
\begin{equation}
S_{\Lambda _x} \lbrack Q \rbrack = {\pi \over 8}N_3(0)D_x 
\int d^3r \Bigl \lbrack {\rm Tr} \lbrack \bbox{ 
\nabla }_\parallel  Q({\bf r})\rbrack ^2 +(d\Lambda _x)^2 {\rm Tr} \lbrack 
\nabla _z Q({\bf r})\rbrack ^2 \Bigr \rbrack 
- {\pi \over 2}N_3(0) \int d^3r
{\rm Tr}\lbrack \Omega Q({\bf r})\rbrack  \,,
\label{action3}
\end{equation}
where $N_3(0)=N_2(0)/d$ is the 3D density of states at the Fermi level. The 
cut-offs are $\Lambda _x$ and $1/d$ in the longitudinal and transverse
directions, respectively. Note that ${\bf r}$ is now a 3D coordinate. 
The 3D regime is thus described by an anisotropic NL$\sigma $M. 
However, the anisotropy is the same for the
diffusion coefficients and the cut-offs and can therefore easily
be suppressed by an appropriate rescaling of the lengths: ${\bf
r}'_\parallel = {\bf r}_\parallel /s_1$, $z'=z/s_2$, with 
\begin{eqnarray}
s_1^{-2}D_x&=&s_2^{-2}D_x(\Lambda _xd)^2 \,, \nonumber \\
s_1^2s_2&=&1 \,.
\label{rescale}
\end{eqnarray}
The last equation ensures that $\int d^3{\bf r}'=\int d^3{\bf r}$ and lets
invariant the last term of (\ref{action3}). From (\ref{rescale}), we obtain
$s_1=(1/\Lambda _xd)^{1/3}$, $s_2=(\Lambda _xd)^{2/3}$, and the new
effective action
\begin{equation}
S_{\Lambda _x} \lbrack Q \rbrack = {\pi \over 8}N_3(0)\bar D 
\int d^3r {\rm Tr} \lbrack \bbox{ \nabla } Q({\bf r})\rbrack ^2 
- {\pi \over 2}N_3(0) \int d^3r
{\rm Tr}\lbrack \Omega Q({\bf r})\rbrack  \,,
\label{action4}
\end{equation}
where $\bar D=D_x(\Lambda _xd)^{2/3}$. The cut-off is now isotropic:
$s_1\Lambda _x\sim s_2/d \sim \bar \Lambda =(\Lambda _x^2/d)^{1/3}$. The
dimensionless coupling constant of the NL$\sigma $M (\ref{action4}) is 
\begin{equation}
\lambda =\frac{4}{\pi N_3(0)\bar D} \bar \Lambda = \frac{4}{\pi N_2(0)D_x}
=\frac{4}{\pi g_x} \,. 
\end{equation}
The MIT occurs for $\lambda =\lambda _c=O(1)$, i.e. when the
2D conductance $g_x=g_c=O(1)$. Using $g_x=g_c\sim 1$
for $\Lambda _x\sim \xi _{2D}^{-1}$, we recover the result (\ref{tcrit}) for
the critical value of the interplane coupling. 

\begin{figure}
\epsfysize 7cm
\epsffile[50 250 450 500]{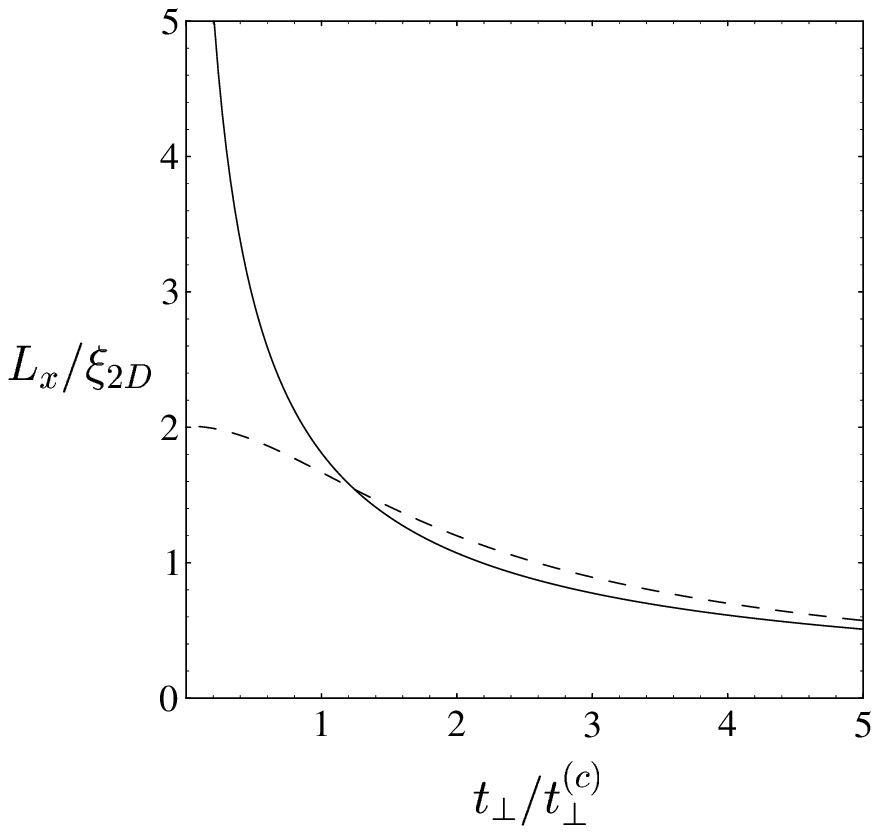}
\caption{Crossover length $L_x$ vs interplane coupling $t_\perp $. Solid
line: RG approach, dashed line: auxiliary field method. We have used
$k_Fl=50$ and $\xi _{2D}/l=5\,10^4$ ($l$ and $\xi _{2D}$ were considered as
independent parameters).    }  
\label{FigLX}
\end{figure}

Further information about the metallic and insulating phases can be obtained
from the $L$-dependence of the 2D dimensionless conductance. The
self-consistent diagrammatic theory of Anderson localization
gives\cite{Vollhardt92} 
\begin{equation}
g(L)={1 \over {2\pi ^2}} \ln \left ( 1+\frac{\xi _{2D}^2}{L^2} \right )
\left ( 1+\frac{L}{\xi_{2D}} \right ) e^{-L/\xi _{2D}} \,,
\label{g2D}
\end{equation}   
where $L$ should be identified with $\Lambda ^{-1}$. 
Eq.\ (\ref{g2D}) gives $g(L)\simeq (1/\pi ^2)\ln (\xi _{2D}/L)$ for
$L\ll  \xi _{2D}$ (in agreement with perturbative RG
calculation\cite{Belitz94}) and $g(L)\simeq   
(1/2\pi ^2)(\xi _{2D}/L)e^{-L/\xi _{2D}}$ for $L\gg  \xi _{2D}$. The
crossover length $L_x$ obtained from (\ref{critXover}) and (\ref{g2D}) is
shown in Fig.\ \ref{FigLX}. Since the crossover
length is not precisely defined, we have multiplied $L_x$ by a constant in
order to have $L_x\simeq v_F/t_\perp $ deep in the metallic phase ($t_\perp \gg
t_\perp ^{(c)}$) (this will allow a detailed comparison with the results of
Sec.\ \ref{sec:AF}).     

\begin{figure}
\epsfysize 10cm
\epsffile[80 235 450 525]{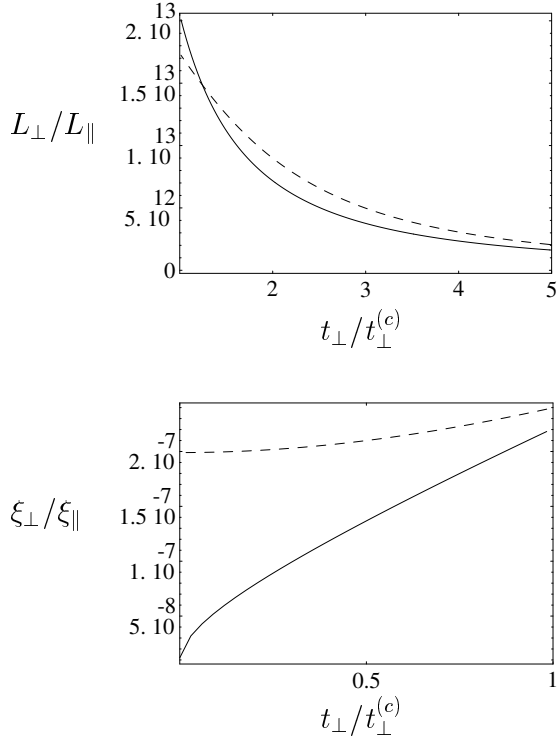}
\caption{Anisotropy of the correlation (localization) lengths for $t_\perp
>t_\perp ^{(c)}$ ($t_\perp < t_\perp ^{(c)}$). Solid line: RG approach,
dashed line: auxiliary field method. The parameters are the same as in Fig.\
\ref{FigLX} with $k_Fd\sim 1$. }   
\label{FigAnis}
\end{figure}

In the metallic phase, $t_\perp \geq t_\perp ^{(c)}$, we have
\begin{eqnarray}
g_x &\simeq & 2N_2(0)t_\perp ^2 \tau \xi _{2D}^2 e^{-2\pi ^2g_x} \gtrsim 1\,, 
\nonumber \\
L_x &\simeq & \xi _{2D} e^{-\pi ^2g_x} \lesssim \xi _{2D} \,.
\end{eqnarray}
Since the renormalization in the 3D regime does not change the anisotropy,
we obtain from (\ref{action3}) the anisotropy of the correlation lengths
(see Fig.\ \ref{FigAnis}):
\begin{equation}
\frac{L_\perp }{L_\parallel }= \frac{\sigma _\parallel }{\sigma _\perp }
=\frac{1}{(\Lambda _xd)^2} \,.
\end{equation}
Deep in the metallic phase ($t_\perp \gg t_\perp ^{(c)}$), we have $L_\perp
/L_\parallel \sim (t_\parallel /t_\perp )^2$ (where $t_\parallel \sim
v_Fk_F\sim v_F/d$ is the transfer integral in the planes) while close to the
transition ($t_\perp \gtrsim t_\perp ^{(c)}$) we have $L_\perp /L_\parallel
\sim (\xi _{2D}/d)^2\sim (t_\parallel /t_\perp )^2/(k_Fl)$. Thus, as we move
towards the transition, the anisotropy deviates from the result $L_\perp
/L_\parallel \sim (t_\parallel /t_\perp )^2$ predicted by the 3D anisotropic
NL$\sigma $M. \cite{Wolfle84}

In the insulating phase, $t_\perp \leq t_\perp ^{(c)}$, we have 
\begin{eqnarray}
g_x &=& \frac{1}{2\pi ^2}\xi _{2D} \sqrt{\frac{2N_2(0)t_\perp ^2\tau }{g_x}}
e^{ -\sqrt{\frac{g_x}{2N_2(0)t_\perp ^2\tau }} \frac{1}{\xi _{2D}}} \lesssim
1 \,,  \nonumber \\
L_x &\simeq & \sqrt{\frac{g_x}{2N_2(0)t_\perp ^2\tau }} \gtrsim \xi _{2D}
\,.
\end{eqnarray}
The anisotropy of the localization lengths is given by (see Fig.\
\ref{FigAnis}) 
\begin{equation}
\frac{\xi _\perp }{\xi _\parallel }=\frac{s_2}{s_1}=\Lambda _xd \,.
\end{equation}
Close to the transition, $\Lambda _x\sim \xi _{2D}^{-1}$, so that $\xi
_\perp /\xi _\parallel \sim \sqrt{k_Fl}(t_\perp /t_\parallel )$. Again this
differs from the result $\xi _\perp /\xi _\parallel \sim t_\perp
/t_\parallel $ predicted by a 3D anisotropic NL$\sigma $M.\cite{Wolfle84}

\subsection{Effect of a parallel magnetic field} 
\label{subsec:MF}

\begin{figure}
\epsfysize 6cm
\epsffile[-200 80 650 700]{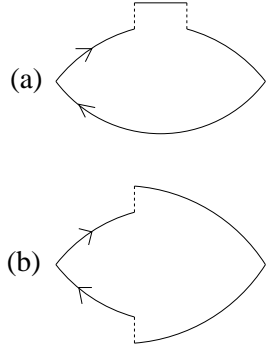}
\caption{Diagrammatic representation of $\Pi ^{(1)}$ (the symmetric diagram
is not shown) (a) and $\Pi ^{(2)}$ (b). 
The solid lines represent the 2D Green's functions while the dashed
lines denote single particle interplane hopping. } 
\label{FigDia}
\end{figure}

We consider in this section the effect of an external magnetic field ${\bf
H}=(0,H,0)$ parallel to the planes. In second quantized form, the interplane
hopping Hamiltonian is 
\begin{eqnarray}
{\cal H}_\perp &=& -t_\perp \sum _{\langle l,l'\rangle } \int d^2{\bf r}\,
e^{ie\int _{({\bf r},ld)}^{({\bf r},l'd)} {\bf A}({\bf s}) \cdot d{\bf s} } 
\psi ^\dagger ({\bf r},l) \psi ({\bf r},l') \nonumber \\
&=& -t_\perp \sum _{\langle l,l'\rangle ,{\bf k}_\parallel } \psi ^\dagger
({\bf k}_\parallel +(l-l'){\bf G},l') \psi ({\bf k}_\parallel ,l)
\label{Hperp}
\end{eqnarray}
in the gauge ${\bf A}(0,0,-Hx)$. ${\bf G}=(G,0,0)$ with $G=-eHd$.
In the second line of (\ref{Hperp}), we have 
used a mixed representation by taking the Fourier transform with respect to
the intraplane coordinate ${\bf r}$. The effective action $S\lbrack Q\rbrack
$ of the NL$\sigma $M can be simply obtained by calculating the
particle-hole bubble $\Pi ({\bf q},\omega _\nu )$ in the semiclassical
(diffusive) approximation. To lowest order in $t_\perp $, we have $\Pi =\Pi
^{(0)}+\Pi ^{(1)}+\Pi 
^{(2)}$ where $\Pi ^{(0)}\simeq 2\pi N_2(0)\tau (1-\vert \omega _\nu \vert
\tau -D\tau q_\parallel ^2)$ (for $\vert \omega _\nu \vert \tau ,D\tau
q_\parallel ^2\ll 1$) is the 2D result. Here $\omega _\nu $ is a bosonic
Matsubara frequency. $\Pi ^{(1)}$ and $\Pi ^{(2)}$ are
given by (see Fig.\ \ref{FigDia}) 
\begin{eqnarray}
\Pi ^{(1)} &=& \frac{2t_\perp ^2}{L^2} \sum _{{\bf k}_\parallel ,\delta
=\pm 1} G^2({\bf k}_\parallel ,\omega _n) 
G({\bf k}_\parallel -\delta {\bf G},\omega _n) 
G({\bf k}_\parallel ,\omega _{n+\nu }) \nonumber \\
&=& -8\pi N_2(0)\frac{t_\perp ^2\tau ^3}{(1+\omega _c^2\tau
^2)^{1/2}} \,, \nonumber \\
\Pi ^{(2)} &=& \frac{t_\perp ^2}{L^2} \sum _{{\bf k}_\parallel ,\delta
=\pm 1} e^{iq_\perp \delta d} 
G({\bf k}_\parallel ,\omega _n) 
G({\bf k}_\parallel -\delta {\bf G},\omega _n) 
G({\bf k}_\parallel ,\omega _{n+\nu })
G({\bf k}_\parallel -\delta {\bf G},\omega _{n+\nu }) \nonumber \\
&=& 8\pi N_2(0)\frac{t_\perp ^2\tau ^3}{(1+\omega _c^2\tau
^2)^{1/2}}\cos (q_\perp d) \,,
\end{eqnarray}
for $\omega _n\omega _{n+\nu }<0$ and $q_\parallel ,\omega _\nu \to
0$. $L^2$ is  the area of the planes and
\begin{equation}
G({\bf k}_\parallel ,\omega _n)=(i\omega
_n+(i/2\tau ){\rm sgn}(\omega _n)-{\bf k}_\parallel ^2/2m+\mu )^{-1}
\end{equation}
is the 2D one-particle Green's function ($\mu $ is the Fermi energy). 
We have introduced the characteristic
magnetic energy $\omega _c=v_FG$. The diffusion modes are determined by
\begin{equation}
1-(2\pi N_2(0)\tau )^{-1} \Pi ({\bf q},\omega _\nu )= \vert \omega _\nu
\vert \tau +D\tau q_\parallel ^2 + \frac{8t_\perp ^2\tau ^2}{(1+\omega _c^2\tau
^2)^{1/2}} \sin ^2(q_\perp d/2) \,,
\end{equation}
which yields the following interplane coupling in the NL$\sigma $M:
\begin{equation}
S_\perp \lbrack Q\rbrack = -\frac{\pi }{4} N_2(0) \frac{t_\perp ^2\tau }
{(1+\omega _c^2\tau ^2)^{1/2}} \sum _{\langle l,l'\rangle } \int d^2r 
\sum _{inm\alpha \beta } {_{ii}Q}_{lnm}^{\alpha \beta }({\bf r}) 
_{ii}Q_{l'mn}^{\beta \alpha }({\bf r}) \,. 
\label{SperpH}
\end{equation}
Notice that we have retained only the diagonal part of the field $Q$ since the
magnetic field suppresses the interplane diffusion modes in the
particle-particle channel. This is strictly correct only above
a characteristic field corresponding to a ``complete'' breakdown of time
reversal symmetry (this point is further discussed below). Eq.\
(\ref{SperpH}) shows that the magnetic field not only breaks down time
reversal symmetry but also reduces the amplitude of the interplane
hopping. This quantum effect (important only when $\omega _c\tau \gg
1$) can be understood from the consideration of the semiclassical electronic
orbits.\cite{ND94} Notice that such an effect cannot be described in the
semiclassical phase integral (or eikonal) approximation for the magnetic
field. The action $S_{2D}$ of the independent planes is
not modified by the magnetic field since the latter is parallel to the
planes. 

We can now apply the same RG procedure as in the preceding section. The
initial (2D) stage of the renormalization is not modified by the parallel
magnetic field, and the 2D/3D dimensional crossover is determined by
\begin{equation}
\frac{1}{2}D_x\Lambda _x^2\sim \frac{t_\perp ^2\tau }{(1+\omega _c^2\tau
^2)^{1/2}} \,.
\label{LxH}
\end{equation} 
The crossover length $L_x(t_\perp ,\omega _c)$ therefore satisfies the
scaling law
\begin{eqnarray}
L_x (t_\perp ,\omega _c)&=& L_x \left ( \frac{t_\perp }{(1+\omega _c^2\tau
^2)^{1/4}},0 \right ) \nonumber \\
&\equiv & L_x \left ( \frac{t_\perp }{(1+\omega _c^2\tau
^2)^{1/4}} \right ) \,,
\end{eqnarray}
where $L_x(t_\perp )$ is the zero field crossover length obtained in the
preceding section. 

In the 3D regime, the diffusion modes are 3D. The smallest volume
corresponding to a
diffusive motion is of order $L_x^2d$. This corresponds to a magnetic flux
$HL_xd$ for a parallel field. Time reversal symmetry is ``completely''
broken when this magnetic flux is at least of the order of the flux quantum,
i.e. when $L_x \gtrsim 1/G$. It is easy to verify that this defines a
characteristic field $H_0$ such that $\omega _c\tau \ll 1$. Above $H_0$, the
diffusion modes are completely suppressed in the particle-particle channel.
The effective action is then given by (\ref{action3})
where only the diagonal part ($i=j$) of the field $Q$ should be
considered. $D_x$ and $\Lambda _x$ depend on $H$ according to
(\ref{LxH}).

\begin{figure}
\epsfysize 8cm
\epsffile[60 250 450 500]{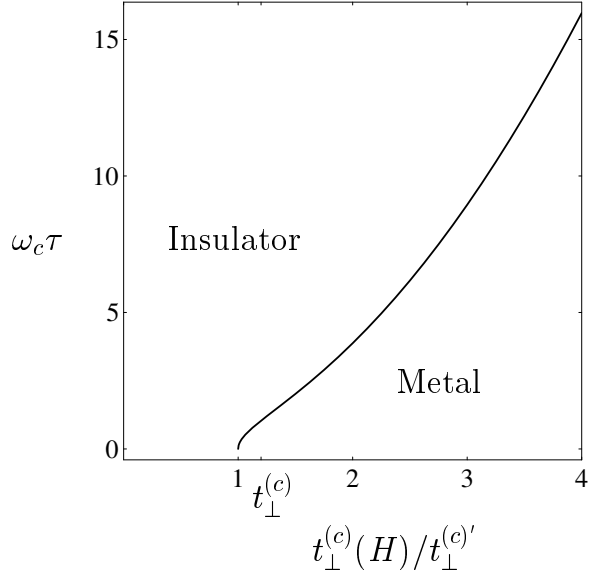}
\caption{Phase diagram in a parallel magnetic field. The position of
$t_\perp ^{(c)}$ ($>t_\perp ^{(c)'}$) is arbitrary. At very low field,
$\omega _c\tau \sim l/\xi _{2D}$,
the curve is not correct since one should have $t_\perp ^{(c)}(H=0)=t_\perp
^{(c)}$ (see text).  } 
\label{FigPhase}
\end{figure}

The critical value of the interplane coupling for $H>H_0$ 
is obtained from $g_x=g_c'$
where $g_c'$ is the critical value of the dimensionless conductance in the
unitary case (no time-reversal symmetry). Thus, we have
\begin{equation}
t_\perp ^{(c)}(H)= {t_\perp ^{(c)}}' (1+\omega _c^2\tau ^2)^{1/4} \,,
\end{equation}
where ${t_\perp ^{(c)}}'\sim \lbrack g_c'/(\tau N_2(0)\xi _{2D}^2)\rbrack
^{1/2}< t_\perp ^{(c)}$ is the critical
coupling for $\omega _c=0$ in the unitary case. Close to the MIT, we have 
$L_x\sim \xi _{2D}$ (since $g_c'\sim 1$) so that
$H_0$ is defined by $\omega _c\sim v_F/\xi _{2D}$, which corresponds to an
exponentially small value of the field: $\omega _c\sim \tau ^{-1}e^{-\alpha
k_Fl}$. The phase diagram in the $(t_\perp -\omega _c)$ plane is
shown in Fig.\ \ref{FigPhase}. For $H\lesssim H_0$, the curve is not correct
and should reach $t_\perp ^{(c)}$ at $H=0$. Therefore, we obtain that a weak
magnetic field favors the metallic phase ($t_\perp ^{(c)}(H)< t_\perp
^{(c)}$ for $0<H\ll H_0$) while a strong magnetic field favors the
insulating phase ($t_\perp ^{(c)}(H)> t_\perp ^{(c)}$ for $H\gg H_0$). This
agrees with the results of Ref.\ \onlinecite{ND92} obtained from scaling
arguments based on the weak localization correction to the Drude-Boltzmann
conductivity. \cite{notaND92}

In the metallic phase, the anisotropy of the correlation
lengths is given by
\begin{equation}
\frac{L_\perp }{L_\parallel } = L_x^2  \left ( \frac{t_\perp }{(1+\omega
_c^2\tau ^2)^{1/4}} \right ) \frac{1}{d^2} \,.
\end{equation}
In the insulating phase, the anisotropy of the localization lengths is given
by
\begin{equation}
\frac{\xi _\perp }{\xi _\parallel } = \frac{d}{ L_x \left ( \frac{t_\perp
}{(1+\omega _c^2\tau ^2)^{1/4}} \right )} \,.
\end{equation}

\section{Auxiliary field method}
\label{sec:AF}

The aim of this section is to recover the results of the RG approach by
means of a completely different method. In order to study how the
``correlations'' are able to propagate in the transverse direction, we will
study an effective field theory which is generated by a Hubbard
Stratonovitch transformation of the interplane coupling $S_\perp \lbrack Q
\rbrack $. This methods bears some obvious similarities with standard
approaches in critical phenomena.\cite{Ising} It is particularly useful when
the low dimensional problem (here the 2D Anderson localization) can be
solved (at least approximately) by one or another method. This kind of
approach has already been used for weakly coupled 1D Ginzburg-Landau models
\cite{McKenzie} or weakly coupled Luttinger liquids. \cite{Boies95} While
these studies were done in the high temperature (or disordered) phase, we
start in this paper from the ``ordered'' (i.e. metallic) phase where the
low-energy modes are Goldstone modes. 

Introducing an auxiliary matrix field $\zeta $ to decouple $S_\perp
$, we rewrite the partition function as 
\begin{equation}
Z= \int {\cal D}\zeta \,
e^{ -\sum _{l,l'} \int d^2r {\rm Tr}\lbrack \zeta _l({\bf r}) J_{\perp
l,l'}^{-1}  
\zeta _{l'}({\bf r})\rbrack } \int {\cal D}Q \,e^{-S_{2D}\lbrack Q\rbrack 
+2 \sum _l \int d^2r {\rm Tr} \lbrack \zeta _l({\bf
r})Q_l({\bf r})\rbrack } \,.
\end{equation}
The field $\zeta $ should have the same structure as the field $Q$ and
therefore satisfies the condition $\zeta ^+=C^T\zeta ^TC=\zeta $. $J_{\perp
l,l'}^{-1}$ is the inverse of the matrix $J_{\perp l,l'}=J_\perp (\delta
_{l,l'+1}+\delta _{l,l'-1})$ where $J_\perp =(\pi /4)N_2(0)t_\perp ^2\tau $
(the matrix $J_\perp $ is diagonal in the indices $\alpha ,n,i$).  

We first determine the value of the auxiliary field in the saddle point
approximation. Assuming a 
solution of the form $_{ij}(\zeta ^{\rm SP})_{lnm}^{\alpha \beta }({\bf
r})=\zeta _0\delta _{i,j}\delta _{\alpha ,\beta }\delta _{n,m}$, we obtain the
saddle point equation
\begin{equation}
\zeta _0=J_\perp (q_\perp =0)\left \langle _{ii}Q_{lnn}^{\alpha \alpha
}({\bf r}) \right \rangle _{\rm SP} \,,
\label{SPeq}
\end{equation}
where $J_\perp (q_\perp )=2J_\perp \cos (q_\perp d)$ is the Fourier transform
of $J_{\perp l,l'}$. The average $\langle \cdot \cdot \cdot \rangle _{\rm
SP}$ should be taken with the saddle point action
\begin{equation}
S_{\rm SP}\lbrack Q\rbrack = S_{2D}\lbrack Q\rbrack-2\sum _l\int d^2r {\rm
Tr} \lbrack \zeta ^{\rm SP}Q_l({\bf r})\rbrack  \,.
\end{equation}
Note that the saddle point value $\zeta ^{\rm SP}$ acts as a finite external
frequency. The mean value of the field Q is related to the density of states
and is a non-singular quantity. We have\cite{Belitz94} $\langle
_{ii}Q_{lnn}^{\alpha \alpha }
({\bf r}) \rangle _{\rm SP}={\rm sgn}(\omega _n)$ which yields 
\begin{equation}
\zeta _0=J_\perp (q_\perp =0){\rm sgn}(\omega _n)=2J_\perp {\rm sgn}
(\omega _n) \,.   
\label{zeta0}
\end{equation}    
This defines a characteristic frequency for the 2D/3D dimensional crossover 
\begin{equation}
\omega _x=\frac{8}{\pi N_2(0)} \vert \zeta _0\vert =4t_\perp ^2\tau \,.  
\end{equation} 

We now consider the fluctuations around the saddle point solution. As in the
standard localization problem, the Sp(2$N$) symmetry of the Lagrangian is
spontaneously broken by the ``frequency'' $\Omega $ to ${\rm Sp}(N)\times
{\rm Sp}(N)$. \cite{Belitz94} The term ${\rm
Tr} \lbrack \Omega Q\rbrack $ breaks the symmetry of $S_{2D}\lbrack Q\rbrack
$. Via the coupling ${\rm Tr}\lbrack \zeta Q\rbrack $ between the fields
$\zeta $ and $Q$, it also breaks the symmetry of the effective action
$S\lbrack \zeta \rbrack $ of the field $\zeta $. Our aim is now to
obtain the effective action of the (diffusive) Goldstone modes associated
with this spontaneous symmetry breaking. Following
Ref.\ \onlinecite{Belitz94}, we shift the field according to $\zeta \to \zeta +
\zeta ^{\rm SP}-\pi N_2(0)\Omega /4$ and expand the action to lowest order
in $\Omega $ and $\zeta $. The partition function becomes 
\begin{eqnarray}
Z &=& \int {\cal D} \zeta \,e^{ -\sum _{l,l'}\int d^2r \bigl \lbrack 
{\rm Tr}\lbrack \zeta _l({\bf r}) J_{\perp l,l'}^{-1} \zeta _{l'}({\bf r})
\rbrack  
+2 {\rm Tr}\lbrack \zeta ^{\rm SP} J_{l,l'}^{-1} \zeta _{l'}({\bf r})\rbrack
-(\pi /2)N_2(0){\rm Tr}\lbrack \Omega J_{l,l'}^{-1} \zeta _{l'}({\bf
r})\rbrack \bigr \rbrack }  
\nonumber \\ && \times \int {\cal D}Q\,
e^{-\tilde S_{2D}\lbrack Q\rbrack +2\sum _l \int d^2r 
{\rm Tr}\lbrack \zeta _l({\bf r}) Q _l({\bf r})\rbrack } \,,
\end{eqnarray}
where we have introduce the 2D action
\begin{eqnarray}
\tilde S_{2D}\lbrack Q\rbrack &=&S_{2D}\lbrack Q\rbrack +\frac{\pi }{2}N_2(0)
\sum _l\int d^2r {\rm Tr}\lbrack \Omega Q_l({\bf r})\rbrack  
-2\sum _l \int d^2r {\rm Tr}\lbrack \zeta ^{\rm SP} Q_l({\bf r})\rbrack 
\nonumber \\ &=&
\frac{\pi }{8}N_2(0)D\sum _l \int d^2r {\rm Tr}\lbrack 
\bbox{ \nabla }_\parallel Q_l({\bf r})\rbrack ^2 
-2\sum _l \int d^2r {\rm Tr}\lbrack \zeta ^{\rm SP} Q_l({\bf r})\rbrack \,.
\end{eqnarray}
$\tilde S_{2D}$ is the action of the decoupled planes ($t_\perp =0$) at the
finite frequency $\omega _x$. To proceed further, we note that 
\begin{equation}
\int {\cal D}Q\,
e^{-\tilde S_{2D}\lbrack Q\rbrack +2\sum _l \int d^2r 
{\rm Tr}\lbrack \zeta _l({\bf r}) Q _l({\bf r})\rbrack } = \tilde Z_{2D}
e^{W\lbrack \zeta \rbrack } \,,
\end{equation}
where $W\lbrack \zeta \rbrack $ is the generating functional of connected
Green's functions calculated with the action $\tilde S_{2D}$.\cite{Negele} 
$\tilde Z_{2D}$ is the partition function corresponding to the action 
$\tilde S_{2D}$. We have  
\begin{eqnarray}
W\lbrack \zeta \rbrack &=& 2\sum _l\int d^2r {\rm Tr}\lbrack \zeta _l({\bf r})
\langle Q_l({\bf r})\rangle _{\tilde S_{2D}}\rbrack  \nonumber \\ &&
+2\sum _l \int d^2r_1d^2r_2  
\sum _{ijnm\alpha \beta } {_{ij}\zeta }_{lnm}^{\alpha \beta }({\bf r}_1) 
_{ij}\tilde R_{lnm}^{\alpha \beta }({\bf r}_1,{\bf r}_2) 
_{ji}\zeta _{lmn}^{\beta \alpha }({\bf r}_2) +... 
\end{eqnarray}
where\cite{nota1}
\begin{equation}
_{ij}\tilde R_{lnm}^{\alpha \beta }({\bf r}_1,{\bf r}_2) = \langle 
_{ji}Q _{lmn}^{\beta \alpha }({\bf r}_1)
_{ij}Q _{lnm}^{\alpha \beta }({\bf r}_2) \rangle _{\tilde S_{2D}} \,.
\end{equation}
Using the saddle point equation (\ref{SPeq}), 
we obtain to quadratic order in the field $\zeta $ and lowest order in 
$\Omega $ the effective action 
\begin{eqnarray}
S\lbrack \zeta \rbrack &=& \sum _{l,l'} \int d^2r_1d^2r_2\,    
\sum _{ijnm\alpha \beta }
{_{ij}\zeta }_{lnm}^{\alpha \beta }({\bf r}_1) 
\Bigl \lbrack J_{\perp l,l'}^{-1} \delta ({\bf r}_1-{\bf r}_2) 
-2 {_{ij}\tilde R}_{lnm}^{\alpha \beta }({\bf r}_1,{\bf r}_2) \delta _{l,l'} 
\Bigr \rbrack \,
_{ji}\zeta _{lmn}^{\beta \alpha }({\bf r}_2) \nonumber \\ &&
-\frac{\pi }{2}N_2(0)\sum _{l,l'} \int d^2r {\rm Tr}\lbrack 
\Omega J_{\perp l,l'}^{-1} \zeta _l({\bf r})\rbrack  \,.
\end{eqnarray}
For $\omega _n\omega _m<0$, $ {_{ij}\tilde R}_{lnm}^{\alpha \beta }({\bf
r}_1,{\bf r}_2)$ is the propagator of the Goldstone modes of the action
$\tilde S_{2D}\lbrack Q\rbrack $. Its Fourier transform is given by
\begin{equation}
 {_{ij}\tilde R}_{lnm}^{\alpha \beta }({\bf q}_\parallel )\equiv 
\tilde R({\bf q}_\parallel )= \frac{4}{\pi N_2(0)(D_xq_\parallel ^2+\omega
_x)} \,,
\end{equation}
where $D_x$ is the exact 2D diffusion coefficient at the finite frequency
$\omega _x$.
Notice that the finite frequency $\omega _x$ gives a mass to the Goldstone
modes. The preceding equation defines the crossover length $L_x=(D_x/\omega
_x)^{1/2}$. Since $\tilde R({\bf q}_\parallel =0)=1/2J_\perp (q_\perp =0)$,
the fluctuations of $\zeta $ around its saddle point value are massless for
$\omega _n\omega _m<0$. On the other hand, it is clear that the fluctuations
are massive for $\omega _n\omega _m>0$. Having identified the Goldstone
modes resulting from the spontaneous symmetry breaking, we now follow the
conventional NL$\sigma $M approach. \cite{Belitz94} We suppress the massive
fluctuations imposing on the field $\zeta $ the constraints $\zeta ^2={\zeta
^{\rm SP}}^2$ and ${\rm Tr}\,\zeta =0$. Rescaling the field $\zeta $ in
order to have $\zeta ^2=\underline 1$ and introducing the Fourier
transformed fields $\zeta ({\bf q})$, we obtain
\begin{equation}
S\lbrack \zeta \rbrack = J_\perp ^2 \sum _{\bf q} (J_\perp ^{-1}(q_\perp )
-2\tilde R({\bf q}_\parallel )) {\rm Tr}\lbrack \zeta ({\bf q}) 
\zeta (-{\bf q}) \rbrack -\frac{\pi }{2}N_2(0) {\rm Tr}\lbrack \Omega \zeta
({\bf q}=0)\rbrack \,.
\end{equation}

In the 3D regime, $q_\parallel \lesssim 1/L_x$ and $q_\perp \lesssim 1/d$,
we can expand $J_\perp ^{-1}(q_\perp )-2\tilde R({\bf q}_\parallel )$ in
lowest order in $q_\parallel $ and $q_\perp $ to obtain
\begin{equation}
S\lbrack \zeta \rbrack = \frac{\pi }{8}N_2(0) \sum _{\bf q} 
(D_xq_\parallel ^2 +2t_\perp ^2d^2\tau q_\perp ^2) {\rm Tr}\lbrack \zeta
({\bf q}) \zeta (-{\bf q}) \rbrack  -\frac{\pi }{2}N_2(0) {\rm Tr}\lbrack
\Omega \zeta ({\bf q}=0)\rbrack \,.
\end{equation}
Going back to real space and taking the continuum limit in the $z$ direction
(which introduces a factor $1/d$), we eventually come to 
\begin{equation}
S\lbrack \zeta \rbrack = \frac{\pi }{8}N_3(0) \int d^3r \Bigl \lbrack D_x
{\rm Tr}\lbrack \bbox{ \nabla }_\parallel \zeta \rbrack ^2 +2t_\perp ^2d^2\tau
{\rm Tr}\lbrack \nabla _z \zeta \rbrack ^2 \Bigr \rbrack -\frac{\pi
}{2}N_3(0) \int d^3r {\rm Tr}\lbrack \Omega \zeta \rbrack \,.
\label{action5}
\end{equation}
The cut-offs are $\Lambda _x=L_x^{-1}$ in the longitudinal directions and
$1/d$ in the transverse direction. 

Eq.\ (\ref{action5}) is similar to the action (\ref{action3}) we have
obtained in the RG approach. The only difference is that the crossover
length is not defined in the same way. In the RG approach, $L_x\sim
D(L_x)/\omega _x$ is defined via the length dependent 2D diffusion
coefficient while the auxiliary field method involves the frequency
dependent 2D diffusion coefficient. We approximate the latter by
\cite{Vollhardt92} 
\begin{equation}
D(\omega _\nu )= \frac{D}{1+\frac{l^2}{\xi _{2D}^2\vert \omega _\nu
\vert \tau }} \,, 
\end{equation}
where $\omega _\nu $ is a bosonic Matsubara frequency. This yields 
\begin{equation}
\Lambda _x^2=L_x^{-2}=\frac{\omega _x}{D}\left ( 1+\frac{l^2}{\xi
_{2D}^2\omega _x\tau } \right ) \,.
\end{equation}
The critical interplane coupling $t_\perp ^{(c)}$ is determined by
$N_2(0)D_x\equiv N_2(0)D(\omega _x)\sim 1$ which leads again to
(\ref{tcrit}). The crossover length $L_x$ is shown in Fig.\ \ref{FigLX}. For
$t_\perp \gtrsim t_\perp ^{(c)}$, there is an agreement with the RG
approach. The reason is that in the weak coupling limit ($g\gtrsim 1$), $D(L)$
and $D(\omega _\nu )$ approximately coincide. \cite{Vollhardt92} Deep in the
insulating phase, this is not the case and the two approaches give different
results: $L_x\gg \xi _{2D}$ in the RG approach  while $L_x\sim \xi _{2D}$ in
the auxiliary field method. This disagreement should not be surprising
since neither one of the two methods is exact. In the RG approach, the
dimensional crossover is treated very crudely since all the 
effects due to $t_\perp $ are neglected in the first (2D) stage of the
renormalization procedure. However, this method should give qualitatively
correct results. In particular, we expect the result $L_x\gg \xi _{2D}$ for
$t_\perp \ll t_\perp ^{(c)}$ to be correct. In the auxiliary field method,
the effective action (\ref{action5}) was obtained by completely neglecting 
the massive modes (more precisely sending their mass to infinity). 
In principle, these latter should be integrated out, which
would lead to a renormalization of the diffusion coefficients appearing in
(\ref{action5}). The comparison with the RG result suggests that this
renormalization is important in the insulating phase.  

The generalization of the preceding results in order to include a parallel
magnetic field is straightforward. The interplane coupling is now given by
(\ref{SperpH}). The auxiliary field $\zeta $ has therefore only a diagonal
part ($i=j$). The saddle point equation yields 
\begin{equation}
_{ii}(\zeta ^{\rm SP})^{\alpha \beta }_{lnm}({\bf r})=\delta _{\alpha ,\beta
}\delta _{n,m}{\rm sgn}(\omega _n) \frac{\pi }{2} N_2(0)
\frac{t_\perp ^2\tau }{(1+\omega _c^2\tau ^2)^{1/2}} \,.
\end{equation}
This defines the crossover frequency 
\begin{equation}
\omega _x=\frac{4t_\perp ^2\tau }{(1+\omega _c^2\tau ^2)^{1/2}} \,.
\end{equation}
In the 3D regime, the massless fluctuations around the saddle point value
are described by the action 
\begin{equation}
S\lbrack \zeta \rbrack = \frac{\pi }{8}N_3(0) \int d^3r \Bigl \lbrack D_x
{\rm Tr}\lbrack \bbox{ \nabla }_\parallel \zeta \rbrack ^2 +
\frac{2t_\perp ^2d^2\tau }{(1+\omega _c^2\tau ^2)^{1/2}}
{\rm Tr}\lbrack \nabla _z \zeta \rbrack ^2 \Bigr \rbrack -\frac{\pi
}{2}N_3(0) \int d^3r {\rm Tr}\lbrack \Omega \zeta \rbrack \,,
\label{action6}
\end{equation}
with the usual constraints on the field $\zeta $. Here
$D_x$ is the coefficient diffusion calculated at the magnetic field
dependent frequency $\omega _x$. Again we recover the result of the RG
approach, the only difference coming from the definitions of $L_x$ and
$\Lambda _x$.

\section{Conclusion}

Using two different methods, we have studied the Anderson MIT in quasi-2D
systems. We have found that the critical value of the single particle
interplane coupling is given by (\ref{tcrit}). Apart from the factor
$1/\sqrt{k_Fl}$, this result agrees with the diagrammatic self-consistent
theory of Anderson localization and with estimates based on the weak
localization correction to the Drude-Boltzmann conductivity.
Nevertheless, it differs from recent numerical calculations according to
which $t_\perp ^{(c)}\sim 1/\sqrt{\tau }$. \cite{Zambetaki96} In the weak
disorder limit ($k_Fl\gg1 $), this latter result seems to us in
contradiction with the scaling theory of Anderson localization since 
the 2D localization length $\xi _{2D}$ is exponentially large with
respect to the mean free path $l$. Because of the latter property, we indeed
expect an exponentially small value of the critical coupling $t_\perp
^{(c)}$ with respect to the elastic scattering rate $1/\tau
$. \cite{Affleck} For a very large 2D localization length $\xi _{2D}$, 
it seems unlikely that the dimensional crossover and the MIT can be studied
with numerical calculations on finite systems. The numerical calculations of
Ref.\ \onlinecite{Zambetaki96} are done in a strong disorder regime
($k_Fl\sim 1$). In this regime, the exponential dependence of
$t_\perp ^{(c)}$ on $k_Fl$ may be easily overlooked. 

We have also studied the anisotropy the correlation (localization) lengths
in the metallic (insulating) phase and shown that it differs from the result
predicted by a 3D anisotropic NL$\sigma $M. The phase diagram in presence of
a magnetic field was also derived: our approach formalizes and extends
previous results obtained for weakly coupled chains. \cite{ND92}

\begin{references}
\bibitem{WL} This estimate is obtained from the criterion $\delta \sigma \sim
\sigma $. In Ref.\ \onlinecite{Abrikosov94}, the correct estimate
$t_\perp ^{(c)}\sim \tau ^{-1}e^{-\alpha k_Fl}$ is obtained only from the
longitudinal correction $\delta \sigma _{xx}=\delta \sigma _{yy}$. Since the
weak localization correction $\delta \sigma _{\mu \mu }/\sigma _{\mu \mu }$ is
independent of $\mu =x,y,z$ (see for instance Ref.\
\onlinecite{Prigodin84}), the same estimate of $t_\perp ^{(c)}$ should be
obtained from $\delta \sigma _{zz}$.
\bibitem{possible} This is possible because the anisotropy is the same for
the diffusion coefficients ($D$ and $D_\perp $) and the ultraviolet cut-offs
($1/\sqrt {D\tau }$ and $1/\sqrt {D_\perp \tau }$). \cite{Wolfle84}
\bibitem{notaND92} Note that in Ref.\ \onlinecite{ND92}, 
the planes themselves are
highly anisotropic with an open Fermi surface. In that case, $(1+\omega
_c^2\tau ^2)^{1/2}$ should be replaced by $1+\omega _c^2\tau ^2$ in
(\ref{SperpH}).
\bibitem{nota1} Since we consider only the two-point Green's function, we do
not need to distinguish between connected and disconnected Green's
functions.
\end{references}
\end{document}